\documentclass[twoside,twocolumn,9pt]{article}
\usepackage{extsizes}
\usepackage[super,sort&compress,comma]{natbib} 
\usepackage[version=3]{mhchem}
\usepackage[left=1.5cm, right=1.5cm, top=1.785cm, bottom=2.0cm]{geometry}
\usepackage{balance}
\usepackage{mathptmx}
\usepackage{sectsty}
\usepackage{graphicx} 
\usepackage{lastpage}
\usepackage[format=plain,justification=justified,singlelinecheck=false,font={stretch=1.125,small,sf},labelfont=bf,labelsep=space]{caption}
\usepackage{float}
\usepackage{fancyhdr}
\usepackage{fnpos}
\usepackage[english]{babel}
\addto{\captionsenglish}{%
  
}
\usepackage{array}
\usepackage{droidsans}
\usepackage{charter}
\usepackage[T1]{fontenc}
\usepackage[usenames,dvipsnames]{xcolor}
\usepackage{setspace}
\usepackage[compact]{titlesec}
\usepackage{hyperref}
\usepackage{makecell}

\usepackage{epstopdf}

\definecolor{cream}{RGB}{222,217,201}

\begin{document}

\pagestyle{fancy}
\thispagestyle{plain}
\fancypagestyle{plain}{
\renewcommand{\headrulewidth}{0pt}
}

\makeFNbottom
\makeatletter
\renewcommand\LARGE{\@setfontsize\LARGE{15pt}{17}}
\renewcommand\Large{\@setfontsize\Large{12pt}{14}}
\renewcommand\large{\@setfontsize\large{10pt}{12}}
\renewcommand\footnotesize{\@setfontsize\footnotesize{7pt}{10}}
\makeatother

\renewcommand{\thefootnote}{\fnsymbol{footnote}}
\renewcommand\footnoterule{\vspace*{1pt}%
\color{cream}\hrule width 3.5in height 0.4pt \color{black}\vspace*{5pt}} 
\setcounter{secnumdepth}{5}

\makeatletter 
\renewcommand\@biblabel[1]{#1}            
\renewcommand\@makefntext[1]%
{\noindent\makebox[0pt][r]{\@thefnmark\,}#1}
\makeatother 
\renewcommand{\figurename}{\small{Fig.}~}
\sectionfont{\sffamily\Large}
\subsectionfont{\normalsize}
\subsubsectionfont{\bf}
\setstretch{1.125} 
\setlength{\skip\footins}{0.8cm}
\setlength{\footnotesep}{0.25cm}
\setlength{\jot}{10pt}
\titlespacing*{\section}{0pt}{4pt}{4pt}
\titlespacing*{\subsection}{0pt}{15pt}{1pt}

\fancyfoot{}
\fancyfoot[LO,RE]{\vspace{-7.1pt}\includegraphics[height=9pt]{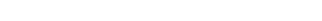}}
\fancyfoot[CO]{\vspace{-7.1pt}\hspace{13.2cm}\includegraphics{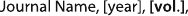}}
\fancyfoot[CE]{\vspace{-7.2pt}\hspace{-14.2cm}\includegraphics{head_foot/RF}}
\fancyfoot[RO]{\footnotesize{\sffamily{1--\pageref{LastPage} ~\textbar  \hspace{2pt}\thepage}}}
\fancyfoot[LE]{\footnotesize{\sffamily{\thepage~\textbar\hspace{3.45cm} 1--\pageref{LastPage}}}}
\fancyhead{}
\renewcommand{\headrulewidth}{0pt} 
\renewcommand{\footrulewidth}{0pt}
\setlength{\arrayrulewidth}{1pt}
\setlength{\columnsep}{6.5mm}
\setlength\bibsep{1pt}

\makeatletter 
\newlength{\figrulesep} 
\setlength{\figrulesep}{0.5\textfloatsep} 

\newcommand{\topfigrule}{\vspace*{-1pt}%
\noindent{\color{cream}\rule[-\figrulesep]{\columnwidth}{1.5pt}} }

\newcommand{\botfigrule}{\vspace*{-2pt}%
\noindent{\color{cream}\rule[\figrulesep]{\columnwidth}{1.5pt}} }

\newcommand{\dblfigrule}{\vspace*{-1pt}%
\noindent{\color{cream}\rule[-\figrulesep]{\textwidth}{1.5pt}} }

\makeatother

\twocolumn[
  \begin{@twocolumnfalse}
{\includegraphics[height=30pt]{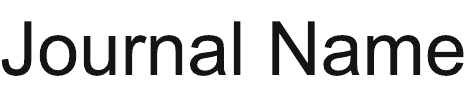}\hfill\raisebox{0pt}[0pt][0pt]{\includegraphics[height=55pt]{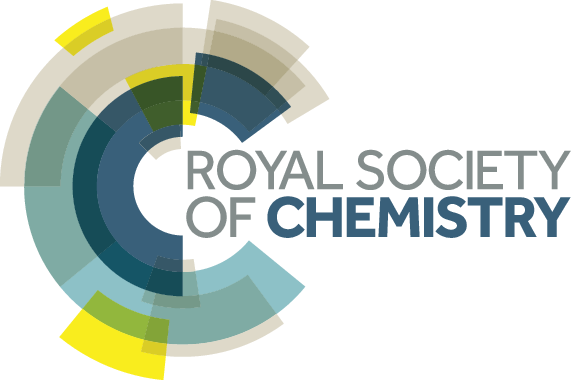}}\\[1ex]
\includegraphics[width=18.5cm]{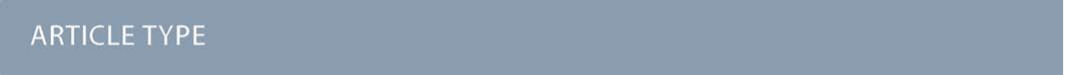}}\par
\vspace{1em}
\sffamily
\begin{tabular}{m{4.5cm} p{13.5cm} }

\includegraphics{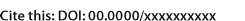} & \noindent\LARGE{\textbf{Efficient Soft-Chemical Synthesis of Large van-der-Waals Crystals of the Room-Temperature Ferromagnet 1T-CrTe\textsubscript{2}$^\dag$}} \\
\vspace{0.3cm} & \vspace{0.3cm} \\

 & \noindent\large{Kai D. Röseler,\textit{$^{a}$} Catherine Witteveen,\textit{$^{a}$} Céline Besnard,\textit{$^{a}$} Vladimir Pomjakushin,\textit{$^{b}$} Harald O. Jeschke,\textit{$^{c}$} and Fabian O. von Rohr$^{\ast}$\textit{$^{a}$}} \\

\includegraphics{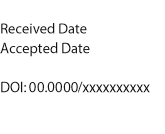} & \noindent\normalsize{We herein report on a fast and convenient soft-chemical synthesis approach towards large 1T-\ce{CrTe2} van-der-Waals crystals. This compound is formed X-ray diffraction pure, with a complete conversion within just over 2\,h from flux-grown \ce{LiCrTe2} crystals using diluted acids. Due to the availability of high-quality single crystals, we have confirmed the crystal structure for the first time by single-crystal X-ray diffraction experiments. For the acid deintercalated 1T-\ce{CrTe2} crystals, we find long-range ferromagnetic order with a Curie temperature of \textit{T}\textsubscript{C}\,=\,318\,K. We further revealed the magnetic structure of 1T-\ce{CrTe2} using low-temperature neutron powder diffraction experiments and determine the magnetic Hamiltonian using density functional theory. X-ray diffraction experiments of post-annealed crystals suggest a thermal stability of 1T-\ce{CrTe2} up to at least 100\,°C. Our findings expand the synthesis methods for 1T-\ce{CrTe2} crystals, which hold promise for integrated room-temperature spintronics applications.} \\

\end{tabular}

 \end{@twocolumnfalse} \vspace{0.6cm}

  ]

\renewcommand*\rmdefault{bch}\normalfont\upshape
\rmfamily
\section*{}
\vspace{-1cm}



\footnotetext{\textit{$^{a}$~Department of Quantum Matter Physics, University of Geneva, 24 Quai Ernest-Ansermet, CH-1211 Geneva, Switzerland}}
\footnotetext{\textit{$^{b}$~Laboratory for Neutron Scattering and Imaging, Paul Scherrer Institute, CH-5232 Villigen PSI, Switzerland}}
\footnotetext{\textit{$^{c}$~Research Institute for Interdisciplinary Science, Okayama University, Okayama 700-8530, Japan}}

\footnotetext{\dag~Supplementary Information available: [details of any supplementary information available should be included here]. See DOI: 00.0000/00000000.}

\footnotetext{$\ast$~To whom correspondence should be addressed.}


\section{Introduction}

Spintronics is an emerging research field to provide the future high-capacity data storage and fast data processing required in information technology.\cite{sierra2021van} Their development is accompanied by the discovery and enhanced synthesis of van-der-Waals (vdW) materials, thereby allowing the fabrication of spintronic devices in the 2D limit.\cite{Ou2022Spin1,Alegria2014_Spin2,Wang2018_Spin3,wu2022quasi} Advanced synthesis methods and especially soft-chemical methods have emerged as indispensable tools to synthesize many of the most promising 2D and van-der-Waals materials for applications.\cite{lotsch2015vertical,lopez2022dynamic,mcqueen2021future,von2017high,song2019soft,sun2015soft}

One of the most promising candidates for 2D spintronic devices is the vdW material 1T-\ce{CrTe2}. Bulk 1T-\ce{CrTe2} has been reported to have a ferromagnetic transition temperature of slightly above room temperature with Curie temperatures ranging between $T_{\rm C}$\,=\,300\,K and 320\,K.\cite{Sun2020,Freitas2015,Purbawati2020,Purbawati2023} Moreover, 1T-\ce{CrTe2} exhibits large magnetic moments, pronounced perpendicular anisotropy, and a spin-split band structure in its magnetic properties.\cite{otero2020controlled,Purbawati2023} Its high Curie temperature has been reported to be nearly retained down to the monolayer level, supported by strong magnetic anisotropy and weak interlayer interactions.\cite{Purbawati2023,Zhang2021} Additionally, 1T-\ce{CrTe2} films were reported to function as efficient spin injectors when combined with other 2D materials like topological insulators and semimetals, facilitating the exploration of new spintronics properties.\cite{Ou2022Spin1,Meng2021,zhang2021giant,fragkos2022magnetic} These characteristics position 1T-\ce{CrTe2} with an exceptional prospect for applications in room-temperature spintronics. The quality of the materials used in these devices is the cornerstone of their further development, which novel synthesis strategies can improve.

\begin{figure*}[h!]
\centering
\includegraphics[width=0.8\textwidth]{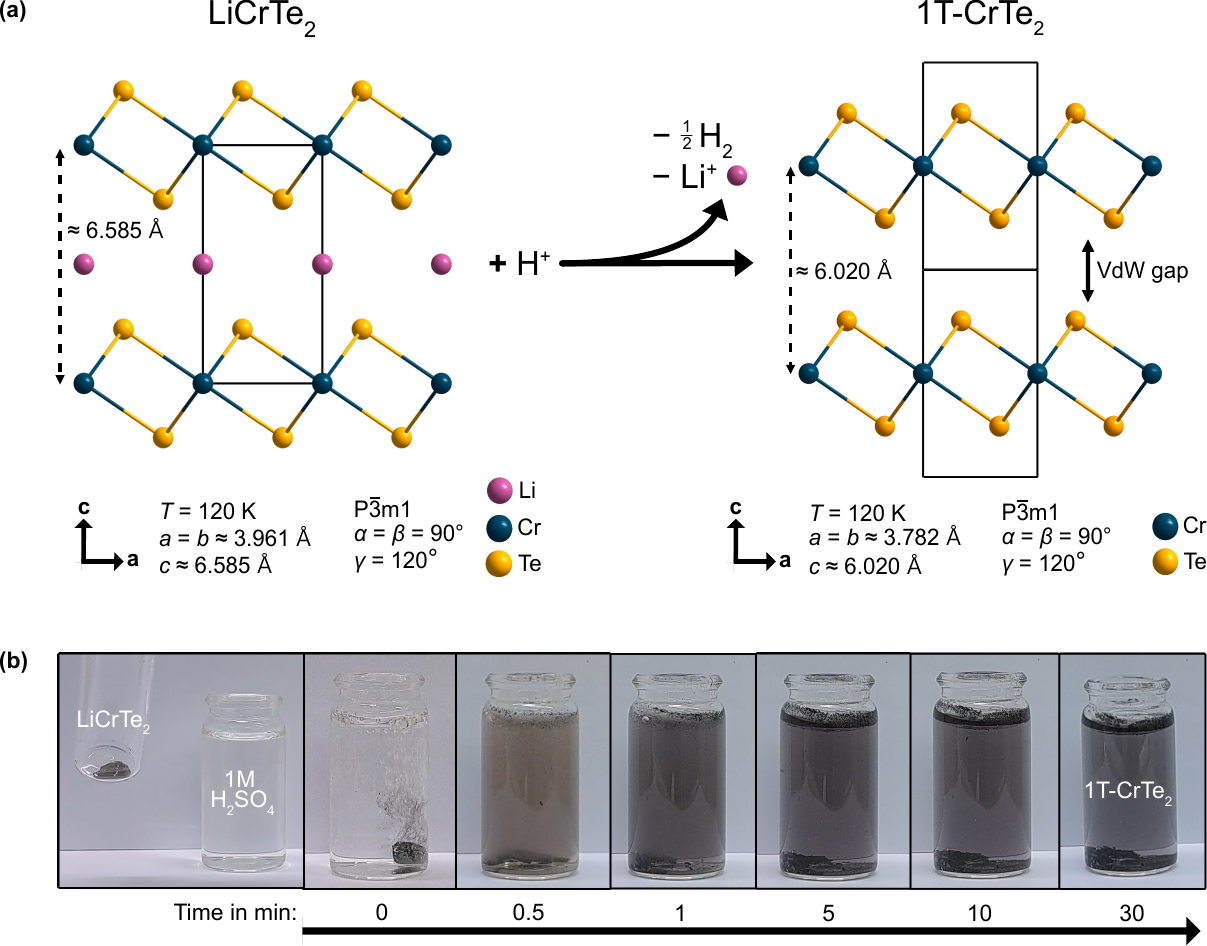}
  \caption{Acid-assisted synthesis of 1T-\ce{CrTe2}: (a) Scheme of the deintercalation reaction of \ce{LiCrTe2} with an acid and the respective cell parameters of \ce{LiCrTe2}\cite{Nocerino2022} based on synchrotron data and 1T-\ce{CrTe2} based on our SXRD data. (b) Images of the deintercalation process of \ce{LiCrTe2} crystals in diluted 1M \ce{H2SO4} with the respective time stamps.}
  \label{fig:flow}
\end{figure*}

Apart from 1T-\ce{CrTe2} other Cr\textsubscript{\textit{x}}Te\textsubscript{\textit{y}} phases have been reported including CrTe\cite{Wang2020CrTe}, \ce{Cr4Te5}\cite{Zhang2020_Cr4Te5}, \ce{Cr3Te4}\cite{Chua2021_Cr3Te4_1,Yamaguchi1972_Cr3Te4}, \ce{Cr2Te3}\cite{Bian2021Cr2Te3_1,Zhong2022_Cr2Te3_2,Hashimoto1971_Cr2Te3_3}, Cr\textsubscript{1+1/3}\ce{Te2}\cite{Lasek2022Cr113Te2,Saha2022_Cr113Te2}, \ce{Cr5Te8}\cite{Bensch_1997_mCr5Te8,Wang2019_Cr5Te8_1,Chen2022_Cr5Te8_2,Tang2022_Cr5Te8_3,Huang2008_Cr5Te8_4}, \ce{Cr3Te5}\cite{Huang2021_Cr3Te5}, and \ce{CrTe3}\cite{Ipser1982_CrTe3}. In contrast to 1T-\ce{CrTe2}, these phases are not vdW materials and thermodynamic products, which can be obtained by reactions of Cr and Te in their respective ratio via solid-state synthesis. These reactions can be exemplary, summarized in phase diagrams such as that by Ipser \textit{et al.}.\cite{Ipser1983}

In addition to this, the synthesis of the meta-stable phase 1T-\ce{CrTe2} has also been reported using two categories of synthesis methods: (i) bottom-up and (ii) top-down approaches. On the one hand, (i) bottom-up syntheses of few-layered or monolayer \ce{CrTe2} have been reported by either using chemical vapor deposition (CVD) directly from the elements on to a substrate\cite{Zhang2021} or molecular beam epitaxy (MBE) from either the elements\cite{Xian2022}, or from \ce{CrCl2} and elemental Te\cite{Meng2021}. Reported (ii) top-down approaches for both crystals and powders, on the other hand, start with the synthesis of the ternary \ce{KCrTe2} from the elements, which in a second step is then deintercalated using \ce{I2} dissolved in acetonitrile\cite{Freitas2015,Purbawati2023,Purbawati2020,Sun2020}.

Employing other soft-chemical deintercalation methods -- as demonstrated exemplarily in the synthesis of the superconductor 2M-\ce{WS2} -- can potentially improve crystallinity and enhance exchange interactions. In the specific case of 2M-\ce{WS2}, Song \textit{et al.} advanced the deintercalation techniques used for K\textsubscript{0.5}\ce{WS2}, from originally involving \ce{K2Cr2O7} and \ce{H2SO4}\cite{Wypych1992,li2021observation,guguchia2019nodeless}, along with reduction using \ce{H2} and subsequent residual deintercalation with \ce{I2}\cite{Lai2021}. They expanded these methods to include complete deintercalation using \ce{I2} and various diluted acids.\cite{Song2023} Hence, expanding the top-down soft-chemical synthesis of vdW materials can play a crucial role in the future fabrication of high-quality quantum materials. 

The growth of single crystals via soft-chemical methods is influenced by the size and quality of the initial crystal, particularly during processes like deintercalation. A recent advance has been the successful growth of large, high-quality \ce{LiCrTe2} crystals using a metal flux composed of Li/Te, which serves as a solvent for Cr.\cite{Witteveen2023} Building on this foundation, we report the synthesis and detailed characterization of large 1T-\ce{CrTe2} crystals synthesized by soft-chemical methods. We employed three different deintercalation techniques: diluted acids, Milli-Q water, and \ce{I2} in acetonitrile, with each being assessed for its impact on the final product. The acid-assisted approach emerged as particularly effective due to its short reaction time of just over two hours, producing large, high-quality 1T-\ce{CrTe2} crystals. This method enabled us to confirm the crystal structure from single-crystal X-ray diffraction data. The resulting acid deintercalated crystals exhibit a ferromagnetic ordering temperature of $T_{\rm C}\,=\,318$\,K. In contrast, deintercalation with \ce{I2}/acetonitrile, requires several days for large crystals to complete, which is significantly longer than the swift acid method. 
Diffraction experiments of post-annealed samples further elucidate the thermal transitions of 1T-\ce{CrTe2}. Neutron diffraction experiments allowed us to solve the magnetic structure of 1T-\ce{CrTe2}. Finally, we used density functional theory to establish a magnetic Hamiltonian for \ce{CrTe2} that clearly confirms our magnetic measurements.

\section{Experimental}

\begin{figure}[h!]
\centering
\includegraphics[width=0.45\textwidth]{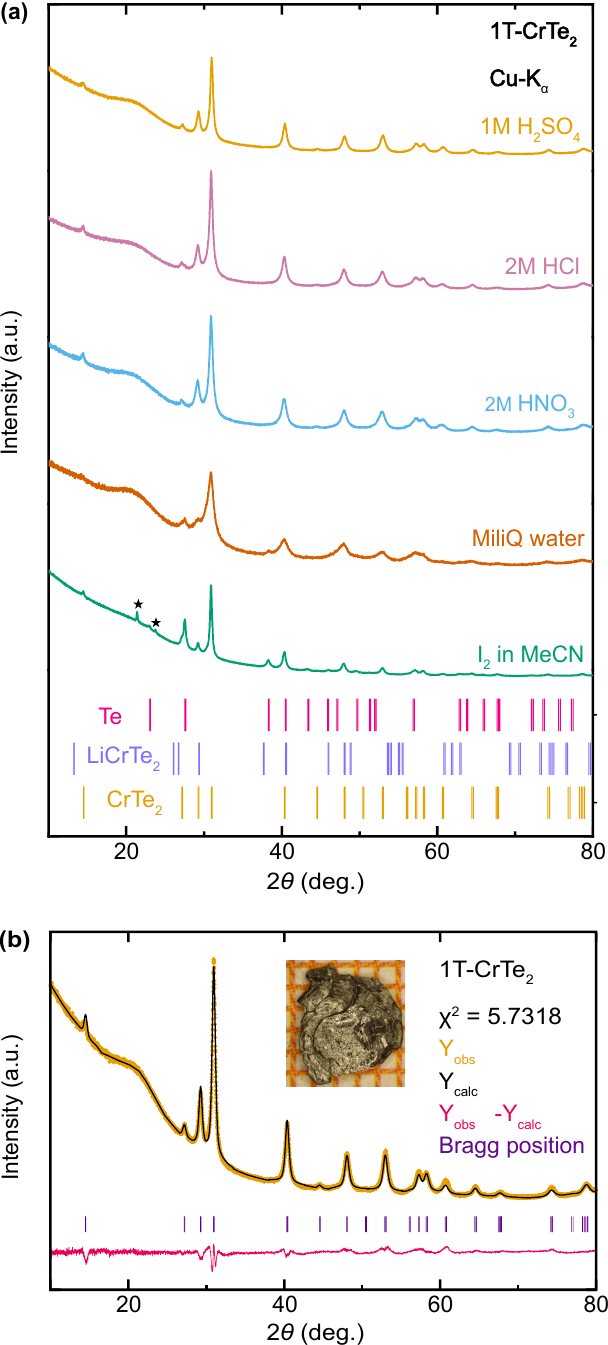}
  \caption{Comparative PXRD patterns of 1T-\ce{CrTe2} synthesized via various deintercalation reactions: (a) PXRD patterns of 1T-\ce{CrTe2} synthesized by deintercaltion of \ce{LiCrTe2} in diluted \ce{H2SO4}, \ce{HCl}, \ce{HNO3}, MiliQ water and \ce{I2} in acetonitrile. The latter pattern was obtained with a capillary measurement instead of reflection mode, which resulted in additional reflections marked with a star, due to Apiezon N Grease used for dilution. A PXRD measurement of the Apiezon N Grease alone in a capillary is depicted in the SI. (b) Rietveld refinement of 1T-\ce{CrTe2} synthesized using a 1M solution of \ce{H2SO4}.}
  \label{fig:PXRD}
\end{figure}

\subsection*{Synthesis}
Single crystals of \ce{LiCrTe2} were synthesized as previously reported from Li (granulates, Sigma-Aldrich, 99\%), Cr (powder, Alfa Aesar, 99.95\%) and Te (pieces, Alfa Aesar, 99.999\%)) using a metal flux method.\cite{Witteveen2023} For the deintercalation in aqueous solutions either 1M \ce{H2SO4}, 2M HCl, 2M \ce{HNO3} or MiliQ water with a ratio of 10\,mg \ce{LiCrTe2} per 1\,ml of the respective solution were used. The reagents were allowed to react for 30\,min after which the solution was replaced three times with 5\,ml of MiliQ water in intervals of 30\,min. Finally, the crystals were rinsed twice in acetonitrile (Sigma-Aldrich, $\geq$ 99.9\%), which was then removed under reduced pressure. The crystals were transferred into an argon-filled glovebox. For the deintercalation using \ce{I2}, a 0.04\,M solution of \ce{I2} (granulates, Honeywell Fluka, $\geq$ 99.8\%) in dry acetonitrile (Thermo scientific, 99.9\%) was used to yield a ratio of \ce{LiCrTe2} to \ce{I2} of 1:1. The crystals were subsequently washed with dry acetonitrile to remove LiI and excess \ce{I2} and then dried under reduced pressure. All acids used were diluted with MiliQ water from 37\% HCl (Fisher Scientific, laboratory reagent grade), 65\% \ce{HNO3} (carlo erba, for analysis) and 96\% \ce{H2SO4} (carlo erba, for analysis).

\subsection*{Powder X-ray diffraction (PXRD)}
PXRD data were collected using a Rigaku SmartLabXE diffractometer with Cu-K\textsubscript{$\alpha$} radiation ($\lambda$\,=\,1.54187\,\AA) on a D/teX Ultra 250 detector in Bragg Brentano geometry in the 2$\theta$ range of 5° to 80°. Capillary measurements were performed on the same instrument in Debye-Scherrer geometry with Cu-K\textsubscript{$\alpha$} radiation. Crystals were ground into fine powders, mixed with Apiezon N grease, and filled into quartz capillaries with an outer diameter of 0.8\,mm. Powder refinements were conducted using the Rietveld method in the Fullprof Suite package software.\cite{Fullprof}

\subsection*{Single X-ray diffraction (SXRD)} 
Single X-ray diffraction (SXRD) experiments were performed under \ce{N2} cooling at 120\,K on a Rigaku XtaLab Synergy-S diffractometer using Mo K\textsubscript{$\alpha$} radiation ($\lambda$\,=\,0.71072\,\AA). Pre-experiment screenings, data collection, data reduction, and absorption correction were performed using the program suite CrysAlisPro.\cite{Rigaku2015} The crystal structure was solved with the dual space method in SHELXT.\cite{Sheldrick2015XT} The Least square refinement of F\textsuperscript{2} was performed using SHLEXL\cite{Sheldrick2015XL} within the Olex2 crystallography software.\cite{Puschmann2009}

\subsection*{Scanning electron microscopy (SEM) and Energy-dispersive X-ray spectroscopy (EDS)} Electron images were obtained from a JEOL JSM-IT800 Scanning electron microscope with an acceleration voltage of 20\,kV. Energy dispersive X-ray spectroscopy (EDS) data was collected with an X-Max\textsuperscript{N} 80 detector from Oxford Instruments. Stoichimetry calculations are based on ten points on three crystals each.

\subsection*{Magnetization experiments} Magnetization versus temperature and magnetization versus magnetic field measurements were carried out in a Physical Property Measurements in a cryogen-free system (PPMS DynaCool) from Quantum Design equipped with the vibrating sample magnetometer (VSM) option. The measurements were performed in a temperature range of $T$ = 1.8 -- 380\,K in the sweep mode at rates of 5\,K$\cdot$min\textsuperscript{$-1$} and 50\,Oe$\cdot$s\textsuperscript{$-1$} in the range of -9 to 9\,T. Arrot plots were created using magnetization versus field data with a sweeping rate of 40\,Oe$\cdot$s\textsuperscript{$-1$} between 2\,T and 0\,T.

\subsection*{Neutron powder diffraction}
Neutron powder diffraction experiments were performed on the High-Resolution Powder Diffractometer at the Swiss
Spallation Neutron Source from the Paul Scherrer Institute in Villigen, Switzerland.\cite{HRPT} Mortared 1T-\ce{CrTe2} deintercalated from \ce{LiCrTe2} using 1M \ce{H2SO4} was sealed in a vanadium sample container with a diameter of 6\,mm using indium wire in a helium glovebox. Diffraction data were collected at $T\,=\,1.6$\,K with wavelengths of 1.886\, \AA\, and 1.494\, \AA\, as well as at $T\,=\,323$\,K with a wavelength of 1.886\, \AA. The patterns obtained were refined using the Rietveld method using the Fullprof Suite package. The magnetic symmetry was analyzed using ISODISTORT in the ISOTROPY software.\cite{Isodistort,Isodisplace}

\subsection*{Calculations} 

The Hamiltonian of \ce{CrTe2}was determined by density functional theory-based energy mapping.\cite{Jeschke2015,Jeschke2019} We use the all electron  full potential local orbital (FPLO) code~\cite{Koepernik1999} for all density functional theory calculations, in combination with the generalized gradient approximation (GGA) exchange and correlation functional.\cite{Perdew1996}

\subsection*{Post-annealing experiments} 40\,mg of 1T-\ce{CrTe2} synthesized by deintercalation with 1M \ce{H2SO4} was placed in an \ce{Al2O3} crucible and were sealed in quartz ampules under 300\,mbar of Ar. The quartz ampules were placed for 20\,h in preheated ovens at temperatures of 250\,°C, 325\,°C, 400\,°C, and 500\,°C and consequently quenched in air.

\section{Results and Discussion}
\subsection{Acid-Assisted Deintercalation of \ce{LiCrTe2}}

\begin{figure}[h!]
\centering
\includegraphics[width=0.45\textwidth]{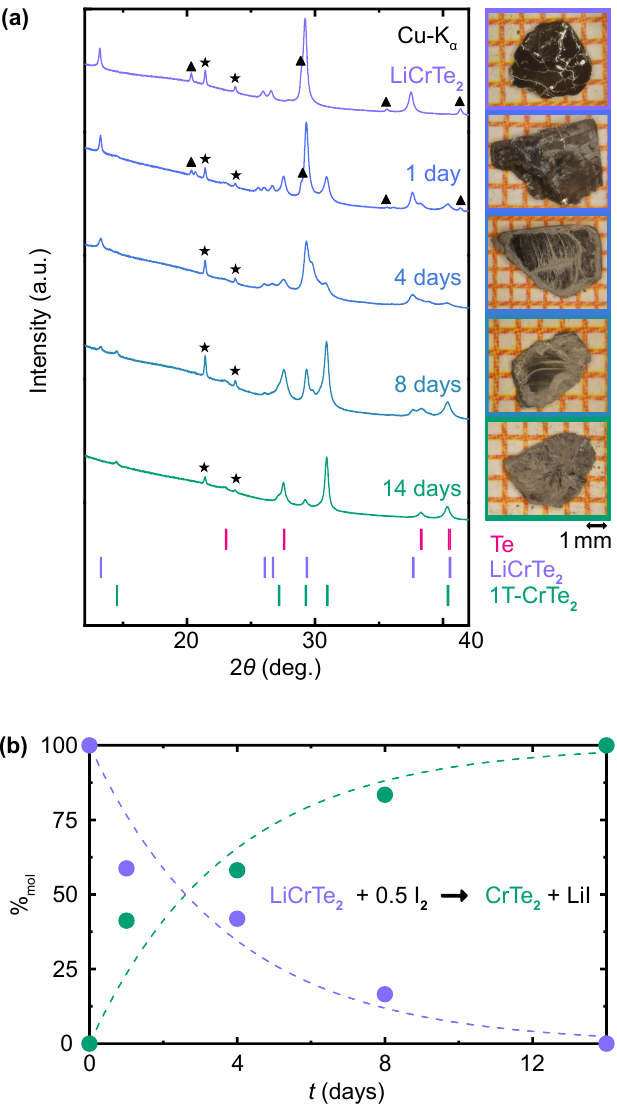}
  \caption{\ce{I2}/acetonitrile deintercalation of \ce{LiCrTe2}: (a) PXRD patterns of \ce{LiCrTe2} and products obtained by deintercalating \ce{LiCrTe2} with a solution of \ce{I2} in acetonitrile for 1, 4, 8, and 14 days. Stars indicate reflections due to Apiezon N Grease used for capillary preparation and triangles reflections due to impurities of \ce{LiTe3}. A PXRD measurement of the Apiezon N Grease alone in a capillary is depicted in the Supplementary Information. Next to the patterns are photographs of the crystals taken on millimeter-sized graph paper. (b) Molar ratios of \ce{LiCrTe2} to 1T-\ce{CrTe2} as a function of time obtained from Rietveld refinements of the patterns above. Dashed lines show a pseudo-exponential fit as guidance. For simplicity, Te was excluded and the sum of molar percentage of \ce{LiCrTe2} and \ce{CrTe2} scaled up to 100\%.}
  \label{fig:PartialDeinter}
\end{figure}

\begin{figure}[h]
\centering
  \includegraphics[width=0.48\textwidth]{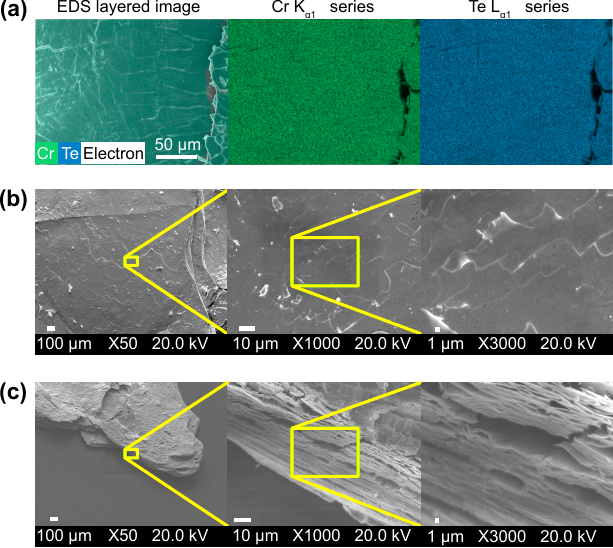}
  \caption{Microscopic Analysis of 1T-\ce{CrTe2} synthesized by deintercalation of \ce{LiCrTe2} in \ce{H2SO4}: EDS map of 1T-\ce{CrTe2} after exfoliation with Scotch tape (a). SEM images of 1T-\ce{CrTe2}. Yellow frames indicate the frame of the respective image with increased magnification. (b) shows images with magnifications of $\times$50, $\times$1000 and $\times$3000 taken perpendicular to the crystal's surface. (c) Shows a side view on a crystal at an angle of about 45° with magnifications of $\times$50, $\times$1000 and $\times$3000.}
  \label{fig:EDSSEM}
\end{figure}

\begin{figure*}
 \centering
 \includegraphics[width=0.8\textwidth]{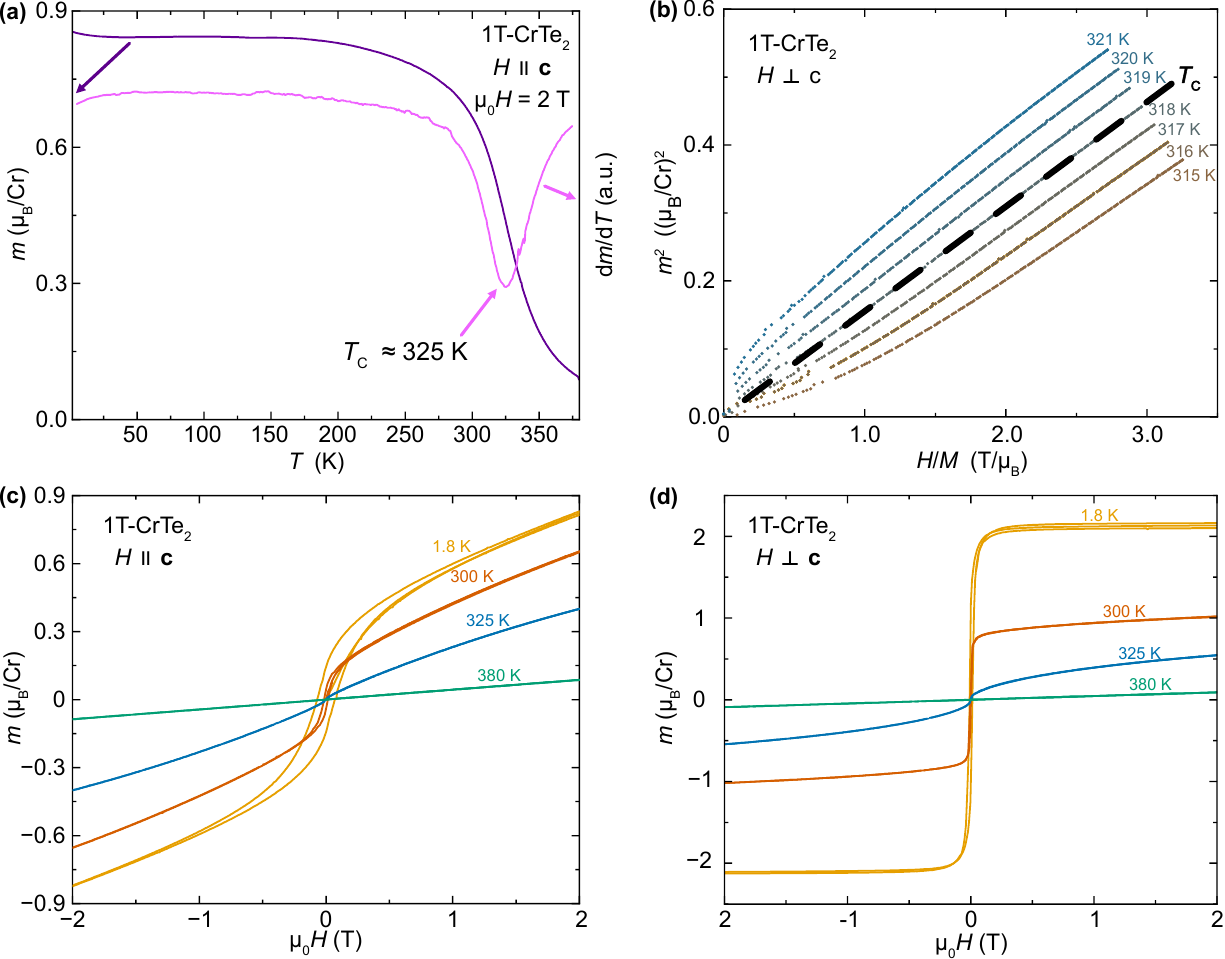}
 \caption{Magnetic properties of \ce{H2SO4}-deintercalated 1T-\ce{CrTe2}: (a) Temperature-dependent magnetization of 1T-\ce{CrTe2} synthesized by deintercalation of \ce{LiCrTe2} in diluted \ce{H2SO4} along the easy axis between 1.8\,K and 380\,K at 2\,T, and its derivative, which was averaged over 10 data points. (b) The ferromagnetic transition temperature was precisely determined at 318\,K using an Arrott plot, (b), which was constructed  from the field-dependent magnetic moment up to 2\,T between 315\,K and 321\,K. Field-dependent magnetic moment between $-2$\,T and 2\,T along the hard axis, (c), and easy plane, (d).}
 \label{fig:magnetic}
\end{figure*}

\begin{figure}[h!]
\centering
  \includegraphics[width=0.45\textwidth]{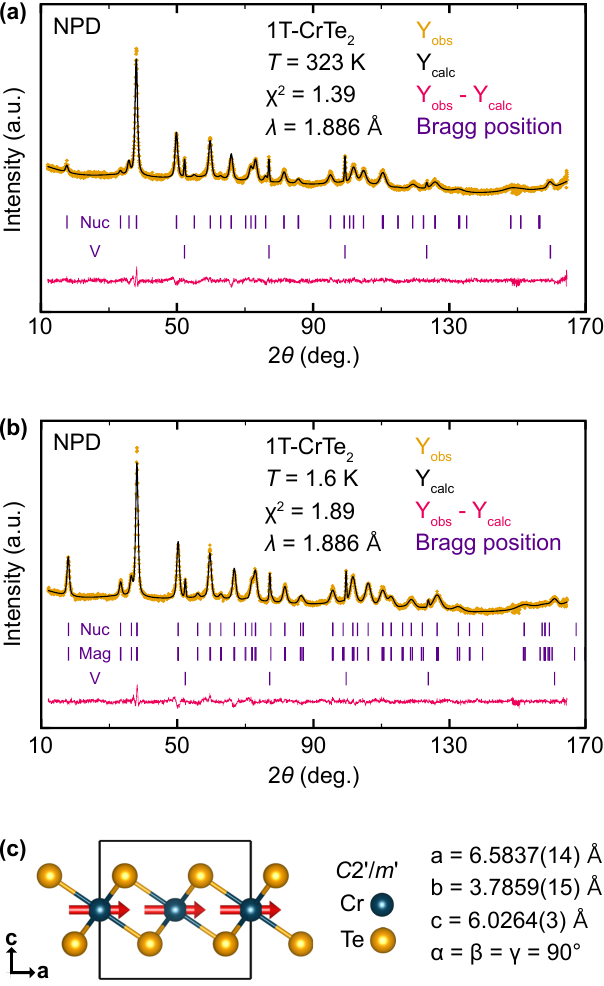}
  \caption{Rietveld refinement of neutron diffraction patterns of \ce{H2SO4}-deintercalated 1T-\ce{CrTe2}. (a) Refined pattern obtained at $T=323$\,K based on two phases: A nuclear part (Nuc) on the basis of SXRD data and the sample container made of vanadium (COD code: 1506411). (b) Refined pattern obtained at $T=1.6$\,K based on three phases: A nuclear part (Nuc) on the basis of SXRD data, a magnetic contribution (Mag) with the space group C2'/m' and the sample contained made of vanadium. (c) Graphical representation of the magnetic structure with parallel orientation of the magnetic moments of Cr represented by red arrows.}
  \label{fig:hrpt}
\end{figure}

\begin{figure}[htb]
\centering
\includegraphics[width=0.4\textwidth]{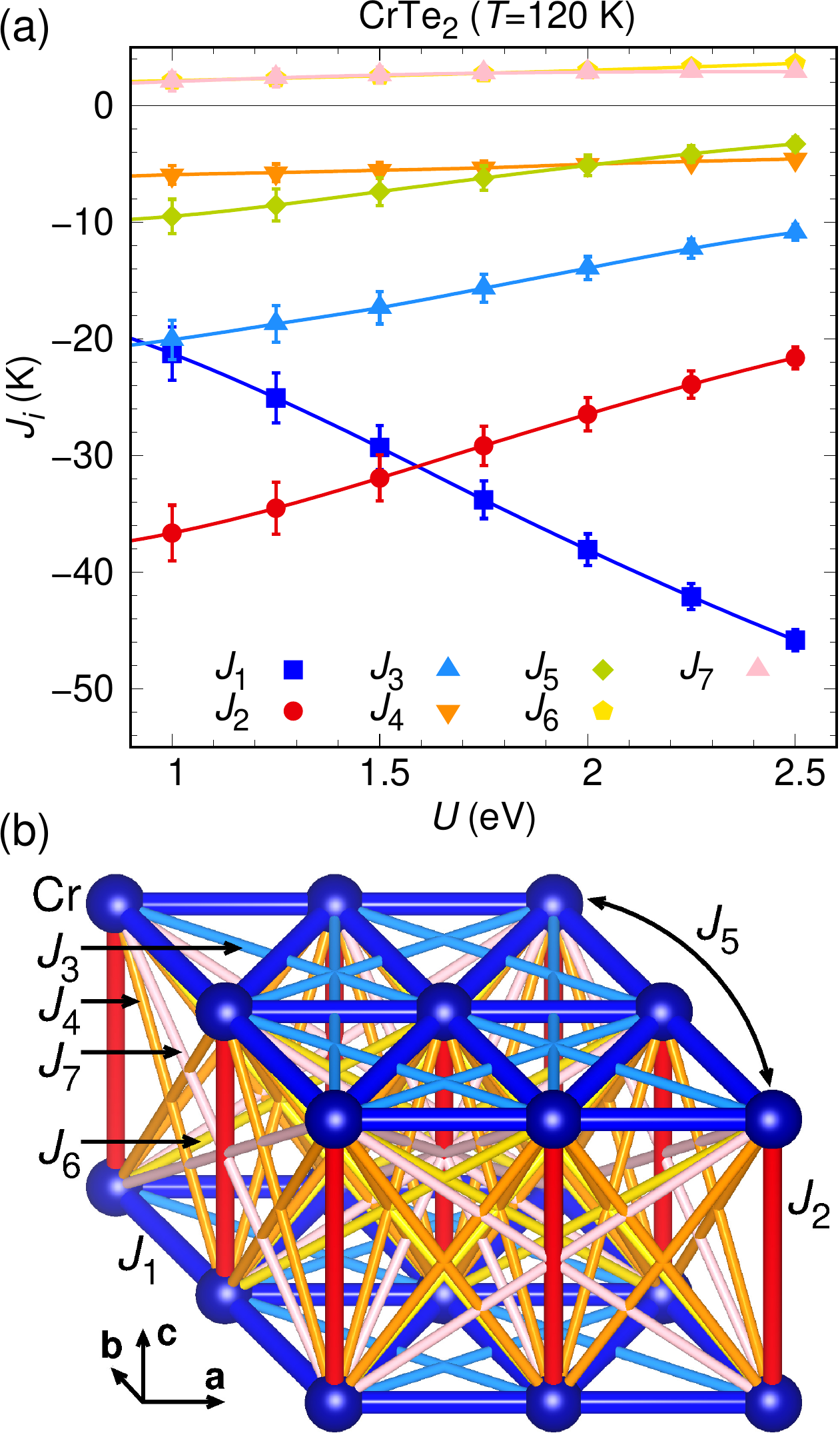}
  \caption{(a) Exchange interactions of 1T-\ce{CrTe2} determined by DFT energy mapping using a DFT+U exchange correlation functional, as function of the on-site interaction strength $U$. (b) Seven relevant exchange paths for \ce{CrTe2}. }
  \label{fig:couplings}
\end{figure}

\begin{figure}[h!]
\centering
\includegraphics[width=0.45\textwidth]{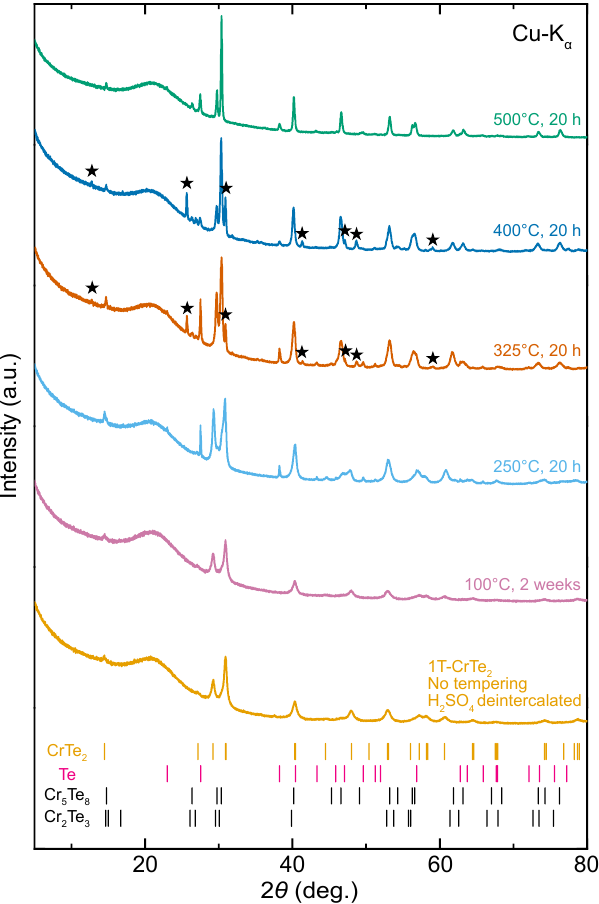}
  \caption{Thermal stability of 1T-\ce{CrTe2}: PXRD patterns of 1T-\ce{CrTe2}, synthesized using diluted acids, tempered at 100\,°C for 2 weeks as well as at 250\,°C, 325\,°C, 400\,°C, 500\,°C for 20\,h. Stars indicate a non-identifiable, compound, possibly with a structure close to \ce{CrTe3}.}
  \label{fig:Temper}
\end{figure}

\noindent In Figure~\ref{fig:flow}\,(a), the schematic of the reaction of \ce{LiCrTe2} with acids are shown. \ce{LiCrTe2} reacts with the acid, i.e. H$^{+}$ ions to result in a redox reaction, in which H\textsuperscript{+} are reduced to \ce{H2}, while the chromium is oxidized from Cr(III) to Cr(IV). The reaction can be followed visually within minutes, as illustrated in Figure~\ref{fig:flow}\,(b) for the case of 1M \ce{H2SO4}.

Visually, the reaction of \ce{LiCrTe2} with diluted \ce{H2SO4}, \ce{HCl}, \ce{HNO3}, and MiliQ water was found to progress in a similar fashion for all of these acids. Upon contact, the formation of the \ce{H2} gas was observed, ending after approximately 5\,min. When washing the crystals with MiliQ water after 30\,min no further formation of gas was observed except in the case of \ce{HNO3}, which then ended again within 5\,min. The second generation of gas when using \ce{HNO3} could indicate an incomplete deintercalation. During the deintercalation process, all diluted acid solution first turned slightly purple but, after being transferred to a separate vial, then became clear within approximately 24\,h.

All synthesized crystals using diluted acids are attracted by a neodymium permanent magnet once it is brought into proximity at room-temperature. This is in agreement with the previously reported room-temperature ferromagnetism for 1T-\ce{CrTe2}. The diameter of the crystals obtained was equal or close to the initial \ce{LiCrTe2} crystals. The size of the \ce{LiCrTe2} used for deintercalation reached diameters up to 8\,mm, which was the diameter of the crucible used for the synthesis. However, we visually observe more phase boundaries due to the lattice strains induced by the structural distortions of the \ce{LiCrTe2} crystals in the periodic changes of the deintercalated crystal (compare, e.g., reference \citenum{peng2017very}).

Perpendicular to the $c$ axis, the crystals can split during the deintercalation. All the obtained crystals are gray with a metallic luster. The crystals show a pronounced layered appearance due to the underlying layered 2D crystal structure.

PXRD measurements of deintercalated \ce{LiCrTe2} using diluted \ce{H2SO4}, \ce{HCl}, \ce{HNO3}, and, for comparison, MiliQ water and \ce{I2} in acetonitrile (after 14 days, see discussion below) -- are depicted in Figure~\ref{fig:PXRD}. These samples deintercalated using \ce{H2SO4}, \ce{HCl}, \ce{HNO3}, and also \ce{I2}/acetonitrile show comparatively sharp peaks in the PXRD patterns. The reflections in the \ce{H2O}-deintercalated sample are wider, implying less crystallinity, and the intensities of the (00\textit{l}) reflections are slightly off, and we observe Te as a clear impurity phase, indicating the partial decomposition of \ce{LiCrTe2} upon reaction with \ce{H2O}. These findings match the comparative deintercalation attempts in the \ce{KWS2}/2M-\ce{WS2} system, in which different soft-chemical methods also yielded products with similar purity and physical properties.\cite{Song2023} Further analyses have been conducted on the sample deintercalated with \ce{H2SO4}. The properties observed in these measurements are likely very similar to those crystals deintercalated with diluted \ce{HCl} and \ce{HNO3} because of the nearly identical diffraction patterns.

The Rietveld refinement, depicted in Figure~\ref{fig:PXRD}\,(b) was conducted on the PXRD-pattern of the \ce{H2SO4}-deintercalated sample based on the respective SXRD crystal structure. The pattern shows no signs of impurities and was refined with $\chi$\textsuperscript{2}\,=\,5.7318. Our Rietveld-refined cell parameters, namely \textit{a}\,=\,3.7875(2)\,\AA, \textit{c}\,=\,6.10323(6)\,\AA, are in excellent agreement with the refined SXRD unit cell parameters discussed below. The peak positions match the patterns and also match the PXRD patterns of all deintercalation methods, hence they all yielded 1T-\ce{CrTe2} crystals. An exemplary image of a \ce{H2SO4} deintercalated crystal is depicted in Figure~\ref{fig:PXRD}\,(b). Images of deintercalated crystals with the other methods are depicted in the Supplementary Information. 

Overall, we show that acid-assisted deintercalation methods of \ce{LiCrTe2} yield phase-pure 1T-\ce{CrTe2} crystals.

\subsection{Deintercalation using Iodine in Acetonitrile}

Single crystals of 1T-\ce{CrTe2} were also synthesized by the adaptation of the \ce{I2} in acetonitrile pathway for the deintercalation of \ce{LiCrTe2} (compare, e.g. reference \citenum{Freitas2015}). The synthesis of high quality 1T-\ce{CrTe2} crystals by this approach was found to be feasible, however significantly slower and accompanied by the presence of Te. These Te impurities are always observed after the deintercalation with \ce{I2}/acetonitrile, while we do not observe it for the acid deintercalation reaction. The PXRD pattern of flux-grown \ce{LiCrTe2} indicates the presence of small amounts of \ce{LiTe3}, which might react with \ce{I2} to Te. The synthesis of millimeter-sized single crystals took 14 days. Shorter reaction times led to products that were not fully deintercalated. In Figure~\ref{fig:PartialDeinter}\,(a), we show the PXRD patterns of crystals with an approximate size of 5 x 3 x 0.5\,mm, which were deintercalated with a 0.04\,M solution of \ce{I2} in acetonitrile for 1, 4, 8 and 14 days. \ce{LiCrTe2} and 1T-\ce{CrTe2} are best identified in the patterns by the peaks at about 13.2° 2$\theta$ and  29.3° 2$\theta$ for \ce{LiCrTe2} or 14.5° 2$\theta$ and 30.9° 2$\theta$ for \ce{CrTe2}. After 1 day, peaks corresponding to \ce{CrTe2} have much smaller intensities than those of \ce{LiCrTe2}. This is also the case after 4 days. After 8 days, the intensity ratio changes, suggesting more 1T-\ce{CrTe2} than \ce{LiCrTe2} is present in the crystal. Only after a duration of 14 days, does the PXRD pattern show no presence of \ce{LiCrTe2} suggesting the deintercalation process to be complete. Rietveld refinements of all patterns were conducted to estimate the ratio of \ce{LiCrTe2} and 1T-\ce{CrTe2} present in the crystal. The molar percentages are depicted in a graph against deintercalation time in Figure~\ref{fig:PartialDeinter}\,(b). The PXRD experiments quantify the percentage of \ce{LiCrTe2} from 100\,\% to 55\,\% after 1 day, to 40\,\% after 4 days, to 13\,\% after 8 days, and to 0\,\% after 14 days.

Upon examining the photographs of the crystals following different deintercalation durations, one can optically observe a dark-gray center surrounded by a light-gray outer ring (see photographs in Figure~\ref{fig:PartialDeinter}\,(a)). As the deintercalation time increases, the light-gray region expands, correlating with the formation of \ce{CrTe2} as confirmed by the powder X-ray diffraction (PXRD) results. Song \textit{et al.} had found for the deintercalation of polycrystalline \ce{KCrSe2} powder using \ce{I2} in acetonitrile a linear progression with time, and suggested a 0\textsuperscript{th} order kinetics reaction in agreement with the K\textsuperscript{+} diffusing towards the edges during the deintercalation.\cite{Song2021} The course of the molar percentages of 1T-\ce{CrTe2} and \ce{LiCrTe2} observed here for millimeter sized crystals matches instead an exponential increase, which hints towards 1\textsuperscript{st} order kinetics, yet the effect of the crystal defects as well as varying crystal sizes cannot be quantified and could have led to this different assumption in kinetics. 

Overall, it can be certainly stated that the overall time necessary for the deintercalation of large \ce{LiCrTe2} crystals using \ce{I2}/acetonitrile is significantly higher than for the acid-assisted deintercalation.

\subsection{Diffraction and Crystal Structure}

For the first time we were able to confirm the crystal structure of 1T-\ce{CrTe2}, which was proposed based on X-ray powder diffraction data by Freitas \textit{et al.}, using SXRD experiments. SXRD could be performed on crystals synthesized within just over 2\,h using diluted \ce{H2SO4}. The best structural model is found to be in the trigonal space group \textit{P}\=3\textit{m}1. The unit cell parameters at 120\,K were found to be almost identical with \textit{a}\,=\,\textit{b}\,=\,3.7823(3)\,\AA, \textit{c}\,=\,6.0203(5)\,\AA\, for the diluted acid deintercalation. The full crystallographic data is listed in Table~\ref{tbl:crydata}. The crystal structure is depicted in Figure~\ref{fig:flow}\,(a) and shows the characteristic layers of the vdW material. Since the Te atoms are arranged octahedrally around the Cr and the Cr---Cr distances within the layers are identical, the 1T polytype can be attributed to this structure. As expected, the Cr---Cr interlayer distance of 6.0205(5)\,\AA\, in 1T-\ce{CrTe2} clearly decreased compared to about 6.585\,\AA\,in \ce{LiCrTe2}. The unit cell consists of two fully occupied crystallographic sites: The 1\textit{b} Wyckoff position, \textit{x}\,=\,0, \textit{y}\,=\,0, \textit{z}\,=\,1/2, is occupied by Cr whereas Te can be found in the 2d Wyckoff position at x\,=\,2/3, y\,=\,1/3, z\,=\,0.2420(3) as listed in Table~\ref{tbl:positions}.

\begin{table}[h]
\small
  \caption{Refined coordinates and assigned Wyckoff position of Cr and Te in 1T-\ce{CrTe2} based on a SXRD measurement at 120\,K.}
  \label{tbl:positions}
  \begin{tabular*}{0.48\textwidth}{@{\extracolsep{\fill}}llllll}
    \hline
    Atom  & Wyckoff & Occ. & x & y & z  \\
    \hline
    Cr & 1\textit{b} & 1 & 0 & 0 & 1/2\\
    Te & 2\textit{d} & 1 & 2/3 & 1/3 & 0.2420(3) \\
    \hline
  \end{tabular*}
\end{table}

The crystal structure of 1T-\ce{CrTe2} from crystals deintercalated using diluted \ce{H2SO4} was solved with reasonable data reduction and refinement parameters with \textit{R}\textsubscript{int}\,=\,0.0954, \textit{R}\textsubscript{1}=0.0678, and \textit{wR}\textsubscript{R2}=0.1757. The model has a significant residual electron density of +11.08\,e\textsuperscript{$-$} \AA\textsuperscript{$-3$}. However, this is located in such close distance to the heavy Te atoms that no additional atom can be reasonably placed in the vdW gap. Both the absence of electron densities between the vdW layers as well as the significant shorter c-axis correspond to the successful deintercalation of Li. Elevated reduction and refinement parameters are most likely the result of the significant mosaicity. Reconstructions of the \textit{hk}0, \textit{h}0\textit{k} and 0\textit{kl} planes of SXRD datasets on crystals deintercalated with diluted acid and \ce{I2} in acetonitrile are depicted in the supplementary information, showing similar mosaicity.

\begin{table}[h!]
\small
  \caption{Crystallographic data for single-crystals of 1T-\ce{CrTe2} synthesized by deintercalation of \ce{LiCrTe2} with diluted \ce{H2SO4}.}
  \label{tbl:crydata}
  \begin{tabular*}{0.48\textwidth}{@{\extracolsep{\fill}}ll}
    \hline
    \textbf{SXRD Refinement} & \textbf{1T-\ce{CrTe2}}\\
    \hline
    Formula  &  \ce{CrTe2}\\
    CCDC Collection code & 2376663\\
    Structure type & \ce{CdI2}\\
    Mol. wt. (g$\cdot$mol\textsuperscript{$-1$}) & 307.20\\
    Crys. syst. & trigonal\\
    Space group & \textit{P}$\Bar{3}$\textit{m}1(164)\\
    \textit{a} (\AA) & 3.7823(3)\\
    \textit{c} (\AA) & 6.0203(5)\\
    \textit{V} (\AA\textsuperscript{3}) & 74.587(13)\\
    \textit{Z} & 1\\
    \makecell[l]{Calculated density\\ (g$\cdot$ cm\textsuperscript{-1})} & 6.839\\
    Temperature (K) & 120\\
    Diffractometer & Synergy, Dualflex, HyPix-Arc 150\\
    Radiation & Mo-K\textsubscript{$\alpha$}\\
    Crystal color & gray\\
    Crystal description & plate\\
    Crystal size (mm\textsuperscript{3}) & $0.24\times0.19\times0.03$\\
    \makecell[l]{Linear absorption\\ coefficient (mm\textsuperscript{-1})} & 22.658\\
    Scan mode & $\omega$ scan\\
    Recording range $\theta$ (°) & 3.264 -- 40.432\\
    \textit{h} range & $-$5 -- 5 \\
    \textit{k} range & $-$6 -- 6 \\
    \textit{l} range & $-$10 -- 10\\
    \makecell[l]{Nb. of measured\\ reflections} & 5693\\
    &\\
    \textbf{Data reduction} &\\
    Completeness (\%) & 100\\
    \makecell[l]{Nb. of independent\\ reflections} & 208\\
    \textit{R}\textsubscript{int} & 0.0954\\
    Absorption corrections & spherical\\
    Independent reflections & 202\\
    with I $\geq$ 2.0$\sigma$ &\\
    & \\
    \textbf{Refinement} &\\
    \textit{R}\textsubscript{1} (obs / all) (\%) & 0.0688 / 0.0694\\
    \textit{wR}\textsubscript{2} (obs / all) (\%) & 0.1855 / 0.1858\\
    \textit{GooF} & 1.335\\
    \makecell[l]{No. of refined\\ parameters} & 6\\
    \makecell[l]{Difference Fourier\\ residues (e\textsuperscript{-} \AA\textsuperscript{-3})} & --4.232 to +11.618\\
    \hline
  \end{tabular*}
\end{table}

Summarized PXRD analysis shows the successful synthesis of 1T-\ce{CrTe2} using diluted acids and iodine in acetonitrile. The previously suggested crystal structure based on powder refinements has been confirmed using SXRD.

\subsection{Microscopic Analysis}

The stoichiometric ratio of Cr to Te was studied using EDS measurements and was found to be nearly ideal with a ratio of $\rm Cr = 1.000\pm0.010$ to $\rm Te = 1.942\pm0.018$. In Figure~\ref{fig:EDSSEM}\,(a), we show an exemplary EDS map with an even distribution of Cr and Te. After exfoliation with Scotch tape, the stoichiometric ratio did not change. This stoichiometry differs significantly from other reported Cr\textsubscript{\textit{x}}Te\textsubscript{\textit{y}} species, EDS therefore substantiates the successful synthesis of \ce{CrTe2}.

The microstructure of a representative acid deintercalated crystal is depicted in Figure~\ref{fig:EDSSEM}\,(b)\&(c). The top-view (b) shows the surface on different scales of the 1T-\ce{CrTe2} crystal. Large areas, i.e., single crystalline domains, can be observed. These areas are interrupted by cracks. These are expected, and likely even unavoidable, by soft-chemical methods in vdW materials (compare, e.g., references \citenum{peng2017very,chung2022mitigating}). The angle-view (c) highlights the layered nature of the resulting crystal, but also emphasizes the presence of disorder, i.e., turbostratic disorder, as the layers have irregular spacings at the edges. This lamellar crystal habitus matches the layered 2D crystal structure of the vdW material. One possible explanation for the irregular spacings at the edges is that the evolved gas leaves the interlayer space towards the edges of the crystal into the solution, applying a perpendicular force on the 1T-\ce{CrTe2} layers. This irregular spacing is likely connected to the mosaicity observed in the SXRD measurement.

Overall, the microscopic analysis confirms the expected stoichiometric ratio of about 1:2 of Cr to Te of 1T-\ce{CrTe2} and showed the layered nature of the vdW-material.


\subsection{Magnetic Properties of 1T-\ce{CrTe2}}

In Figure~\ref{fig:magnetic}, we show the magnetic properties of 1T-\ce{CrTe2} as-prepared using acid-assisted deintercalation from a 1M solution of \ce{H2SO4}. The temperature-dependent magnetization in an external magnetic field of $\mu_0H\,=\,$ 2\,T is shown in \ref{fig:magnetic}\,(a), which reveals the pronounced transition of 1T-\ce{CrTe2} to a ferromagnetic state above room temperature. The transition temperature was determined from the derivative $dM/dT$ at $T_{\rm C}$\,=\,325 K. To quantify the transition temperature of 1T-\ce{CrTe2} crystals more accurately, an Arrott plot\cite{Arrott1957,lopez2022dynamic,lefevre2022heavy} is utilized, as depicted in Figure~\ref{fig:magnetic}\,(b). The Arrott plot -- resulting from mean field theory for magnetism -- corresponds to a $M^2$ vs. $H$/$M$ measurement at fixed temperatures. It is the measurement procedure to (i) provide evidence for the existence of a ferromagnetic long-range ordered state, as well as (ii) for a precise determination of the Curie temperature $T_{\rm C}$ of a ferromagnet. The $M^2$ vs. $H$/$M$ linear behavior that can be extended to the origin of the coordination system for $T_{\rm C}$\,=\,318\,K corresponds to the Curie temperature. Here, this value for the acid deintercalated 1T-\ce{CrTe2} is in agreement with the first derivative of the magnetization and with earlier reports of Curie temperatures between $T_{\rm C}$\,=\,300 K and 320 K.\cite{Sun2020,Freitas2015,Purbawati2020,Fabre2021,Purbawati2023}

The field-dependent magnetization measurements are presented in Figure~\ref{fig:magnetic}\,(c)\&(d) for $T$\,=\,1.8\,K, 300\,K, 325\,K, and 380\,K with the external magnetic field parallel and perpendicular to the $c$ axis respectively. For the measurements with the field along the $c$ axis, at $T$\,=\,1.8 K we observe a clear ferromagnetic behavior displaying a hysteresis loop. Above the Curie temperature, the field-dependent magnetization is linear, as expected in the paramagnetic state. In this orientation of the crystal there is no saturation of the magnetization observed up to $\mu_0 H$\,=\,9\,T as this corresponds to the hard axis (see SI). 

With the magnetic field perpendicular to the $c$ axis, corresponding to the easy plane, no prominent hysteresis can be observed, yet the magnetization almost saturates above \textit{T}\textsubscript{C}\,=\,318\,K. We estimate a saturation moment of around 2.1\,\textmu\textsubscript{B} by extrapolation that is found above $\mu_0 H$\,=\,9 T. This is in agreement with the theoretical moment of Cr\textsuperscript{4+} of $\approx$ 2.82\,\textmu\textsubscript{B} based on the spin only formula \textmu\,=\,$\sqrt{n(n+2)}$ with $n$ being the number of unpaired electrons. Saturation magnetization, as well as direction of the hard axis, are in agreement with earlier reports on samples of 1T-\ce{CrTe2} from \ce{KCrTe2} using \ce{I2}/acetonitrile.\cite{Freitas2015,Sun2020,Fabre2021,Purbawati2020, Purbawati2023}

The magnetic properties of 1T-\ce{CrTe2} prepared via acid-assisted deintercalation from 1M \ce{H2SO4} demonstrate a transition to a ferromagnetic state above room temperature with a Curie temperature of $T_{\rm C} =$ 318 K, confirmed through temperature-dependent magnetization, Arrott plots, and field-dependent magnetization measurements.

\subsection{Neutron Powder Diffraction}

In Figure~\ref{fig:hrpt}, we show the results of neutron powder diffraction experiments that we have performed on finely ground 1T-\ce{CrTe2} crystals, which were synthesized using diluted 1M \ce{H2SO4}. Experiments have been conducted both at $T =$ 1.6\,K and 323\,K, above $T_{\rm C}$\,=\,318\,K. Above the ferromagnetic transition temperature, we successfully refined the diffraction pattern using only the nuclear component, as shown in Figure~\ref{fig:hrpt}\,(a). At $T =$ 1.6\,K, the obtained pattern is well described by introducing a magnetic phase with contributions to the neutron powder pattern at the same 2$\theta$ values as the nuclear part. Based on the nuclear model, four different magnetic space groups (MSG) can be envisioned using ISODISTORT from the ISOTROPY software\cite{Isodistort,Isodisplace}, which are listed in the SI. The highest symmetric space group $P\bar{3}m$'1 was rejected since it would not allow for intensity of the (00\textit{n}) reflections, whereas we observed strong magnetic contributions to the (001) reflection. 
Considering the remaining space groups, we found the long-range magnetic structure is best described in \textit{C}2'/\textit{m}' with the cell parameters \textit{a}\,=\,6.5829(15)\,\AA, \textit{b}\,=\,3.7869(17)\,\AA, \textit{c}\,=\,6.0262(3)\,\AA\, and $\alpha=\beta=\gamma=90°$. Relative to the nuclear structure, a basis transformation with [(2,1,0),(0,1,0),(0,0,1)] was applied. 

Figure~\ref{fig:hrpt}\,(c) depicts a graphical representation of the resulting magnetic structure. It comprises two occupied atom sites with Te located at (2/3, 0, 0.747) and Cr at (0, 0, 0.5). The single magnetic Cr site has a magnetic moment of $\mu_{\rm Cr}=1.329(14)\,\mu_{\rm B}$, which is comparatively low for Cr(+IV) because of the absence of an applied field. Refinement of the magnetic moment vector has led to two solutions which can be considered equally valid based on the obtained value of $\chi^{2}$. A first solution has a magnetic moment vector with contributions both in $x$-direction ($m_{x}=1.311(14)\,\mu_{\rm B}$) and $z$-direction ($m_{z}=0.56(5)\,\mu_{\rm B}$), was refined with $\chi^{2}$\,=\,1.89 and described in more detail in the SI. A second solution has only a contribution of the magnetic moment in the $x$-direction ($m_{x}\,=\,1.329(14)\,\mu_{\rm B}$) and was refined with $\chi^{2}$\,=\,1.92. The second solution with the magnetic moments aligned in the $ab$-plane is the appropriate model, as it agrees better with the observed strong anisotropy between in-plane and out-of-plane magnetization (Figure \ref{fig:magnetic}(c)\&(d)), but also agrees with our fully relativistic energy calculations of the ferromagnetic spins as function of the quantization axis, where the spins clearly prefer to be in the $ab$ plane over the $c$ axis (see discussion below and SI). 

\begin{table}[htb]
\small
  \caption{Comparison of refined parameters of neutron powder diffraction data on \ce{H2SO4}-deintercalated 1T-\ce{CrTe2} collected at 1.6\.K and 323\,K.}
  \label{tbl:npd}
  \begin{tabular*}{0.48\textwidth}{@{\extracolsep{\fill}}lll}
    \hline
    & 1.6\,K & 323\,K\\
    \hline
    \textit{a} (\AA) & 3.79550(12) & 3.7860(3)\\
    \textit{c} (\AA) & 6.0262(3) & 6.1213(6)\\
    \textit{V} (\AA\textsuperscript{3}) & 75.181(5) & 75.985(10)\\
    $\mu_{Cr}$ $(\mu_{\rm B})$ & 1.329(14) & ---\\
    $R_{p}$ & 1.40 & 1.69\\
    $R_{wp}$ & 1.80 & 2.10\\
    $R_{exp}$ & 1.30 & 1.78\\
    $\chi^{2}$ & 1.92 & 1.39\\
    \hline
  \end{tabular*}
\end{table}

The neutron powder diffraction experiments on 1T-\ce{CrTe2} crystals synthesized using diluted 1M-\ce{H2SO4} reveal that the long-range magnetic structure at 1.6 K is best described by the \textit{C}2'/\textit{m}' space group, with Cr atoms having a magnetic moment of $\mu_{\rm Cr}\approx 1.33,\mu_{\rm B}$.

\subsection{Calculated Magnetic Couplings}

Given the availability of a precise crystalline model of 1T-\ce{CrTe2}, we further investigate the magnetic properties of this compound by determining the Heisenberg Hamiltonian $\mathcal{\hat H} = \sum_{i<j} J_{ij} \mathbf{S}_i \cdot \mathbf{S}_j$ where $\mathbf{S}_i$ are spin operators and $J_{ij}$ are Heisenberg Hamiltonian parameters. The Heisenberg Hamiltonian parameters $J_{ij}$ represent the strength and nature of the exchange interactions between the spins of the Cr ions. These parameters are influenced by the electronic structure and the spatial arrangement of the atoms.

We use the well-established approach of density functional theory (DFT) energy mapping, which has previously yielded excellent results for the related compound \ce{LiCrTe2}~\cite{Witteveen2023} as well as for other chromium magnets~\cite{Ghosh2019,Xu2023}. The method implies that we calculate 40 spin configurations with distinct energies for a $3\times2\times2$ supercell and fit their GGA+U energies with the Heisenberg Hamiltonian. This allows us to resolve the seven exchange couplings shown in Fig.~\ref{fig:couplings} for seven values of the onsite Coulomb interaction $U$. 

Our findings indicate that within the triangular lattice formed by Cr ions in the $ab$ plane, the three exchange interactions, denoted as $J_1$, $J_3$, and $J_5$, are ferromagnetic (negative), suggesting that these interactions favor parallel alignment of neighboring spins. Additionally, the interlayer couplings $J_2$, $J_4$, $J_6$, and $J_7$ are predominantly ferromagnetic. This dominance of ferromagnetic interactions both within the plane and between layers corroborates the experimentally observed ferromagnetic order in \ce{CrTe2}. Hence, our calculations confirm that the magnetic measurements of 1T-\ce{CrTe2} reveal predominantly ferromagnetic in-plane and interlayer exchange interactions.

Interestingly, we find in-plane second ($J_3$) and third neighbor ($J_5$) couplings to be substantial in 1T-\ce{CrTe2}. Thus, the high ordering temperature found in the 1T-\ce{CrTe2} monolayer is probably due both to strong single ion anisotropy as well as important longer range exchange interactions~\cite{Jenkins2022}. Meanwhile, the interlayer exchange in 1T-\ce{CrTe2} is not found to be small in our calculations; this is reasonable because the interlayer Cr-Cr distance of 6.0203~{\AA} is small compared to alkali chromium ditellurides, and there are reasonable Cr-Te-Te-Cr exchange paths. Thus, from our calculations we do not find interlayer exchange to be weak, and we can still give valid reasons why the monolayer displays long-range magnetic order. The details on the DFT energy mapping are presented in the SI.

Our calculations confirm that 1T-\ce{CrTe2} exhibits predominantly ferromagnetic in-plane and interlayer exchange interactions, with significant second and third neighbor couplings contributing to the high ordering temperature, and the interlayer exchange being substantial due to the relatively small Cr-Cr distance and viable Cr-Te-Te-Cr exchange paths.

\subsection{Thermal Decomposition}

To investigate the thermal stability and decomposition of acid-assisted deintercalated 1T-\ce{CrTe2} we post-annealed ground powders of acid deintercalated 1T-\ce{CrTe2} crystals under Ar atmosphere. PXRD patterns of the psot-annealed samples are depicted in Figure~\ref{fig:Temper}. The temperatures investigated were inspired by DSC and DTA experiments, which are enclosed in the SI.

Acid deintercalated 1T-\ce{CrTe2} was found thermally stable at least up to 100\,°C for two weeks, after which no significant change of the PXRD pattern compared to the untempered sample was observed. Tempering at 250\,°C resulted in the appearance of reflections that correspond to elemental Te in addition to 1T-\ce{CrTe2}. The observation can be rationalized with the loss of Te in 1T-\ce{CrTe2} due to its thermal degradation. In the PXRD patterns of the samples annealed at 325\,°C and 400\,°C we observe the formation of phases closely related to a Cr-rich composition, corresponding to the thermodynamic stable phases \ce{Cr2Te3} and \ce{Cr5Te8}. The PXRD pattern can be well explained with these structures present. These two phases differ primarily by the amount of Cr that occupies the vdW gap of 1T-\ce{CrTe2}, and can be interpreted as Te-deficient version of the 1T phase, which is well in agreement with the observed Te loss upon thermal treatment. Based on PXRD alone, the clear identification of these two phases from one another is challenging due to their structural similarity. The corresponding reflections in the PXRD pattern that were annealed at 325\,°C and 400\,°C shift slightly relative to each other and there is a difference in their relative intensities, most prominently at 29.7°\,2$\theta$ and 30.4°\,2$\theta$. Given the continuous loss of Te it is likely that at the phase transition 1 the 1T-\ce{CrTe2} phase decomposes into the Te-poorer \ce{Cr2Te3} phase, which at higher temperatures decomposes into the even slightly Te-poorer \ce{Cr5Te8}.

Due to the large number of (meta)-stable Cr\textsubscript{x}Te\textsubscript{y} with stoichiometries close to 1T-\ce{CrTe2} the thermal decomposition of bulk 1T-\ce{CrTe2} and its products is likely a highly complicated process which should be closer investigated in the future. Nevertheless, our observations quantify the thermal decomposition of bulk 1T-\ce{CrTe2} between 100\,°C and 250\,°C. Henceforth, the annealing of these acid deintercalated 1T-\ce{CrTe2} crystals for device fabrication of any sort might be critical in order not to decompose the phase of interest.

The post-annealing experiments reveal that 1T-\ce{CrTe2} remains stable up to 100 \,°C, begins to decompose with the formation of elemental Te at 250 \,°C, and forms Cr-rich phases \ce{Cr2Te3} and \ce{Cr5Te8} at higher temperatures due to thermal degradation and Te loss.

\section*{Conclusions}

In this work, we have expanded the top-down synthesis strategies to yield large, X-ray-pure single-crystals of the room-temperature ferromagnetic vdW material 1T-\ce{CrTe2}. This includes the soft-chemical deintercalation of flux-grown \ce{LiCrTe2}, which allowed the synthesis of large crystals with diameters up to 8\,mm, using diluted acids as well as a solution of \ce{I2} in acetonitrile. We found the deintercalation of \ce{LiCrTe2} using \ce{I2} to be a slow process, i.e. for millimeter-sized crystals the synthesis took up to 14 days, making the acid-assisted deintercalation an efficient and significantly faster alternative, which only took just over 2\,h for crystals. The quality of these acid deintercalated crystals has been found to be high so that we could -- for the first time -- resolve the crystal structure of 1T-\ce{CrTe2} from SXRD data. 

We find for the acid deintercalated 1T-\ce{CrTe2} well-defined magnetic properties with a ferromagnetic transition temperature of  $T\textsubscript{C}= 318$\,K. We have resolved its long-range magnetic ferromagnetic order, using neutron diffraction, which is best described in the magnetic space group \textit{C}2'/\textit{m}', with the moments aligned in the $ab$-plane. Applying DFT-based energy mapping to \ce{CrTe2}, we determined a Heisenberg Hamiltonian with strong ferromagnetic in-plane couplings, including substantial longer range exchange. We find significant interlayer couplings that are also predominantly ferromagnetic. Besides, we obtain a single ion anisotropy that makes 1T-\ce{CrTe2} strongly easy plane. Our calculations can explain the high Curie temperature of the bulk material and give clues why even the monolayer has a high Curie temperature. Finally, our post-annealing experiments of 1T-\ce{CrTe2} suggest a thermal transition of 1T-\ce{CrTe2} into a Cr-richer phase between 100\,°C and 250\,°C marking an important annealing temperature for possible future device fabrications.

In conclusion, the soft-chemical deintercalation of \ce{LiCrTe2} using diluted acids yields high-quality crystals of 1T-\ce{CrTe2} within a short deintercalation time of 2\,h. Future exfoliation of these crystals can be expected to yield few-layered or monolayered samples, which hold promise for the fabrication of room-temperature spintronic devices. 

\section*{Author contributions}
FvR designed the experiments. KR synthesized the crystals. KR, CW, CB, and VP conducted the experiments. HOJ performed the electronic structure calculations. All authors contributed to the analysis of the data. FvR and KR wrote the manuscript with contributions from all the authors.

\section*{Conflicts of interest}
There are no conflicts to declare.

\section*{Data availability}
The data supporting this article have been included as part of the Supplementary Information.

\section*{Acknowledgements}
The authors thank Enrico Giannini and Radovan Cerny for the helpful discussions, Kerry-Lee Paglia for help with the DSC measurement. Part of this work was performed at the Swiss Spallation Neutron Source (SINQ), Paul Scherrer Institut (PSI), Villigen, Switzerland. This work was supported by the Swiss National Science Foundation under Grant No. PCEFP2\_194183.



\balance


\bibliography{rsc} 
\bibliographystyle{rsc} 
\end{document}